# Spectrally-tunable Dielectric Grating-based Metasurface for Broadband Planar Light Concentration

*Ameen Elikkottil‡[1,2], Mohammed H Tahersima‡[3], MVN Surendra Gupta[1,2], Rishi Maiti[3], Volker J. Sorger[3], Bala Pesala\*[1,2]*

[1] Academy of Scientific and Innovative Research, Chennai India

[2] Council of Scientific and Industrial Research-Central Electronic Engineering Research Institute, Chennai, India

[3] Department of Electrical and Computer Engineering, George Washington University, Washington DC, USA

ABSTRACT: The energy consumption of buildings is increasing at a rapid pace due to urbanization, while net-zero energy buildings powered by renewable sources offer a green and sustainable solution. However, the limited rooftop space available on skyscrapers and multi-story buildings poses a challenge for large-scale integration of solar photovoltaic modules. Conventional photovoltaics solutions such as silicon solar panels block the visible light from entering the building rending them unfeasible to cover all building surfaces. Here, we demonstrate a novel dielectric metasurface (based-on silicon nitride grating) acting as a planar light concentrator. We integrate this functional device onto a window glass transmitting visible light while simultaneously guiding the near infrared portion (NIR) of sunlight to the edges of the glass window where it can be converted to electricity by a small PV module. Utilizing the grating design flexibility, we tune the spectra to enable guiding of the near NIR sunlight portion and realize polarization independence demonstrated using finite difference time domain simulations. Experimentally, we observe about 5.25% of optical guiding efficiency in the NIR region (700-1000 nm), leaving majority of the visible portion to be transmitted for natural room lighting. Integrating the solar cell at the window edge, we find a power conversion efficiency of about 4.2% of NIR light on a prototype of area 25 mm$^2$. We confirm that the majority of the loss is due to the absorption, scattering and fabrication non-uniformity over large area which can be further optimized in future. Such a functional window combining renewable energy generation, with room lighting, and building-envelope reduction could mitigate urban heat-islanding issues of modern cities.

1. Introduction

The development of efficient optical concentrators to deliver a functional and economic solution for renewable power harvesting has focused on solar photovoltaic and solar thermal power applications. Prominent methods are utilizing reflection (parabolic dish [1,2]), refraction (Fresnel lens [3-5]), or a combination of both such as compound parabola and prism [6]. Existing concentrators such as the Fresnel reflector, parabolic dish and light funnel[7] require complex opto-mechanical frame works including moving lenses/mirrors to guide/manipulate the direction of light. However, these designs require sophisticated solar tracking to be economical, which demands bulky and costly systems.

In contrast, advances in planar concentrators such as luminescent concentrators [8] (either using dyes [9,10] or quantum dots [11–13]), holographic concentrators [14–16] and micro-optic concentrators [17–19] demonstrate a low-cost and minimally tracking approach for solar light concentration. Luminescent concentrators consist of dyes or quantum dots deposited on a transparent glass substrate [15], which absorb incident light over certain band of wavelengths and re-radiate in a different band either downshifted or up-shifted from the absorbed radiation and subsequently the light is guided in the substrate to edges. Major challenges are the degradation of dyes, self-absorption and low conversion efficiencies [8,20–23]. Moreover, in luminescent solar concentrators absorbed radiation is emitted in all directions resulting in reduced coupling of emitted light into the guided modes of the glass substrate. In addition, in the current designs [8,24,25], relatively narrow band of the incoming solar radiation is harnessed (FWHM <200 nm).

A planar holographic concentrator is a potential technology to overcome these challenges; this system has shown power generation improvements of about 25% in solar cells [26]. However, holographic concentrators block visible light to an extent due to the placement of solar cells in the direct light path. Micro-optic concentrator is another class of planar concentrators in which a micro lens and a diffuser guide incoming sunlight to waveguide. Micro-optic technology can achieve higher concentration because it utilize multiple lenses to focus light and couple to waveguides with rather low angular acceptance (<2°) [17], thus requiring bulky precise tracking infrastructure. Moreover, this technology results in the formation of hotspots, which accelerates solar cell degradation, [27] and results in thermalization losses in solar cell reducing the conversion efficiency in the absence of wavelength selective designs.

While conceptually promising, there are a variety of challenges in existing optical concentration technologies. Grating based designs are one approach to realize a planar concentrator overcoming most of these challenges [28–30]. Fanglu Lu et.al., proposed High Contrast Grating (HCG) as focusing reflectors and lenses using silicon gratings showing numerical apertures of 0.81 and 0.96 respectively [31]. David Fattal et.al., has proposed an aperiodic HCG metasurface as a planar focusing element [32]. Annett B. Klemm et.al., has proposed two-dimensional HCG metasurface as planar micro-lenses [28], and Mohammad Tahersima et al. a metal-based Fabry-Perot cavity solar cell [33]. Here, we explore subwavelength metasurface, or grating based, planar concentrator in this work. For the remainder of this manuscript we use the terminus metasurface and grating interchangeably. Gratings diffract the incoming light and can be designed to couple the light of desired wavelength to first order transmission with highest efficiency by eliminating zero order transmission/reflection. Subwavelength grating based metasurfaces diffract the light at steeper angles over a broad wavelength range with high efficiency and this ensures guiding of desired wavelength band in the glass substrate. Grating based approach has the potential to eliminate the below bandgap longer wavelengths of sunlight reaching the solar cell which otherwise would not contribute to solar photovoltaic conversion and causes heating the solar cell. The desired spectral selectivity of diffraction gratings can be tuned by optimizing the grating geometry parameters: grating period, duty cycle and thickness for a chosen material and optical index.

Planar light concentrators offer significant benefits when used for building integrated photovoltaics. Modern dwellings, such as skyscrapers, often covered with glass, trap solar radiation, impose additional heat load to the cooling systems, and increase the temperature in the surrounding areas, known as urban heat island. In fact, recent studies claim that UHI contributes to 30% of climate warming [34] and can lead to an annual mean air temperature to rise up to 1-3 °C and in the evening as high as 12 °C [35]. Here, we propose and demonstrate a smart-power-window based on a silicon nitride grating based dielectric metasurface for conversion of NIR light to electricity (Fig. 1). This planar concentrator is a multi-spectral metasurface light filter (Fig. 1(a)). It diffracts NIR radiation at steep angles into the window glass substrate where it is guided towards its edge and is converted to electricity using an edge integrated solar cell, reflects thermally parasitic infrared radiation away from the building which reduces the air conditioning load, while simultaneously transmits the visible light to enter the building providing natural room lighting (Fig. 1(b)&(c)).

Our metasurface design is based on a selectively etched silicon nitride gratings on top of a glass substrate (Fig. 1(a) & 3(a)). Geometric design parameters defining spectral light guiding, reflection and transmission coefficients are the grating period ($\Lambda$), the duty cycle (DC), and the thickness ($t_g$). Optimization enables the incident light of desired wavelengths to be diffracted into single transmission order [29] enabling effective guiding of the light to the window edges.

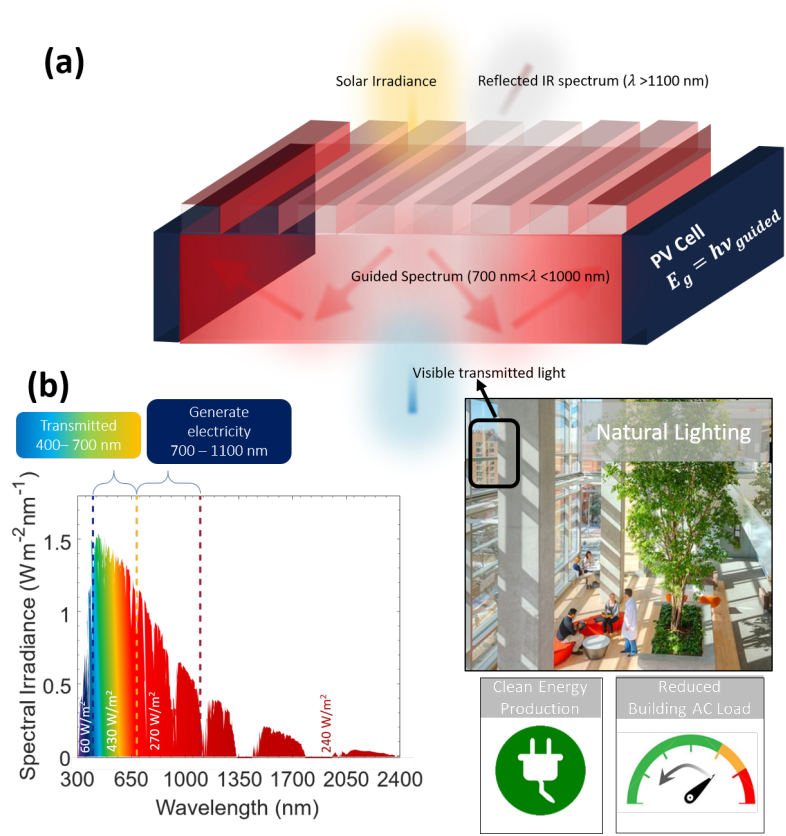

**Figure 1) Proposed Building-integrated photovoltaics system based on a dielectric grating metasurface for broadband spectral filtering forming an optical concentrator**, a) Concept and schematic of the spectrally tunable grating based planar solar concentrator. The grating acts as a spectral filter by diffracting near IR range of the solar radiation at steep angles into the glass substrate. Thus, the PV module at the edge of the glass receives spectrally band-gap

matched radiation, which reduces non-radiative channels and heat. Simultaneously, this setup acts a solar concentrator collecting radiation for energy conversion from the entirety of the grating (e.g. window size), thus reducing the size of PV material needed. Furthermore, the grating material, silicon nitride, is transparent in the visible range, thus enabling natural room lighting. NIR portion of light is converted to electricity, which reduces heat built-up, thus mitigating air conditioning loads. b) The three distinct spectral regions in the sunlight's irradiance are the IR reflection (red), NIR collection (operation range of the grating spectral concentrator, red arrows), and VIS transmission (blue) at AM 1.5G spectrum.

## 2. Results

The metasurface concentrator consists of a transparent glass substrate with high refractive index coating (silicon nitride) followed by a grating of the coating material (Fig. 1(a)&3(a)). The grating is optimized to diffract the NIR portion of the incoming spectrum into the substrate. Therefore, the angular space of the planar concentrator is defined by the grating diffraction condition:

$$\theta_m = sin^{-1}\left(\frac{m\lambda}{\Lambda\, n_{sub}} - \sin(\theta_i)\right) \quad (1)$$

where, $n_{sub}$: refractive index of the substrate, $\theta_i$: the incident angle, $\lambda$: wavelength of the incident light, $\Lambda$: grating period, $m$: diffraction order, and $\theta_m$: the diffraction angle.

Eqn. (1) dictates the grating period to be subwavelength such that the higher diffraction orders contributing to reflection and transmission losses can be minimized and guiding in the desired wavelength range is maximized. Effective guiding in the glass substrate is achieved by exciting first order diffraction modes that are guided inside the glass substrate towards the PV cell. For example, a glass substrate with refractive index of 1.5, grating period of 680 nm, the visible region (400 – 700 nm) diffracts in the angular range of 23.1°– 43.3°. The diffraction angle for NIR wavelength range of 700 – 1000 nm is greater than the critical angle (41.8°) for glass/air interface, hence results in effective guiding into the glass substrate as desired here.

The maximum possible concentration of a HCG-based concentrator for a particular polarization is governed by [30,36]

$$C_{max}(\lambda) \approx \frac{2\, r_d}{1-r_{dt}} \tan(\theta) \quad (2)$$

where, $\theta$ : the diffraction angle for each wavelength, $r_d$: the diffraction efficiency of incoming light to ± 1T, $r_{dt}$: the diffraction efficiency +1T to -1T and vice versa.

However, eqns. 1 and 2 do not provide any insight regarding the diffraction efficiency of the grating structure except the angular distribution of different diffraction orders. Moreover, eqn. 2 does not show the relation of geometric concentration ratio and $C_{max}$. Hence, FDTD simulations are carried out to obtain the optical guiding efficiency and transmission to map-out the geometric parameter space of the grating based light concentrator. We note that these simulations are also carried out with the actual grating profiles obtained from cross-sectional FIB/SEM images to evaluate the performance of fabricated samples.

The choice of materials and geometry directly impacts the grating based metasurface performance. Rigorous Coupled Wave Analysis (RCWA) is used to optimize the grating structure and material. Materials such as silicon, titanium dioxide, silicon dioxide, zinc oxide and silicon nitride on a glass substrate are used in estimating the diffraction efficiencies and angles of the desired wavelength region. RCWA approximation considers the device structure as infinitely periodic but the actual structure is finite in nature, hence we used the Finite Difference Time Domain (FDTD) method instead to evaluate the guiding efficiency (fig. 3). However, FDTD simulation for actual dimension is computationally intensive and hence we have used a scaled down model (15 times) with same geometric ratio. The device structure is optimized with silicon nitride as the grating material since the fabrication process is well developed compared to other materials for our structural dimensions and silicon nitride has a comparatively high refractive index (>2) [37–39], which allows for ease in tuning towards reducing reflection/transmission [40]. In our case, the reflection in the NIR region, should be minimal to reduce optical loss. In addition to this, silicon nitride shows vanishing optical absorption in the visible and NIR [38], has a low sensitivity to temperature variations [41] and is widely used in state-of-the-art CMOS foundries [42]. We optimize the grating for both polarization dependent design (fig. 2(a) & (b)) also for a polarization tolerant design (fig. 2(c)-(e)).

The visible light transmission efficiency ($\eta_{vT}$ -eqn.S1) and guiding efficiency ($\eta_g$ -eqn.S2) are two key metrics used to evaluate the performance of planar concentrators (fig. 2). These efficiencies are sensitive to the grating period, duty cycle and silicon nitride thickness. Visible light transmission accounts for the natural room lighting and guiding efficiency accounts for the generated electricity power from NIR light. One-dimensional grating structures are mostly polarization dependent i.e. the guiding and transmission efficiencies are polarization dependent. Hence, high guiding efficiency can be achieved in one of the polarization, while it will be lower for the other. However, sunlight is unpolarised in nature so it is desired to have a minimal polarization dependence, and hence an overall higher guiding efficiency can be achieved resulting in higher electrical output. Moreover, the guided wavelength and bandwidth can be tuned further to match any solar cell material bandgap towards improving the PV conversion efficiency (i.e. reduced non-radiative recombinations) by adjusting the grating parameters. Here, we consider conventional silicon PV cells and aim for a guided spectral range of 700 – 1000 nm where the Si-PV conversion efficiency is highest [43].

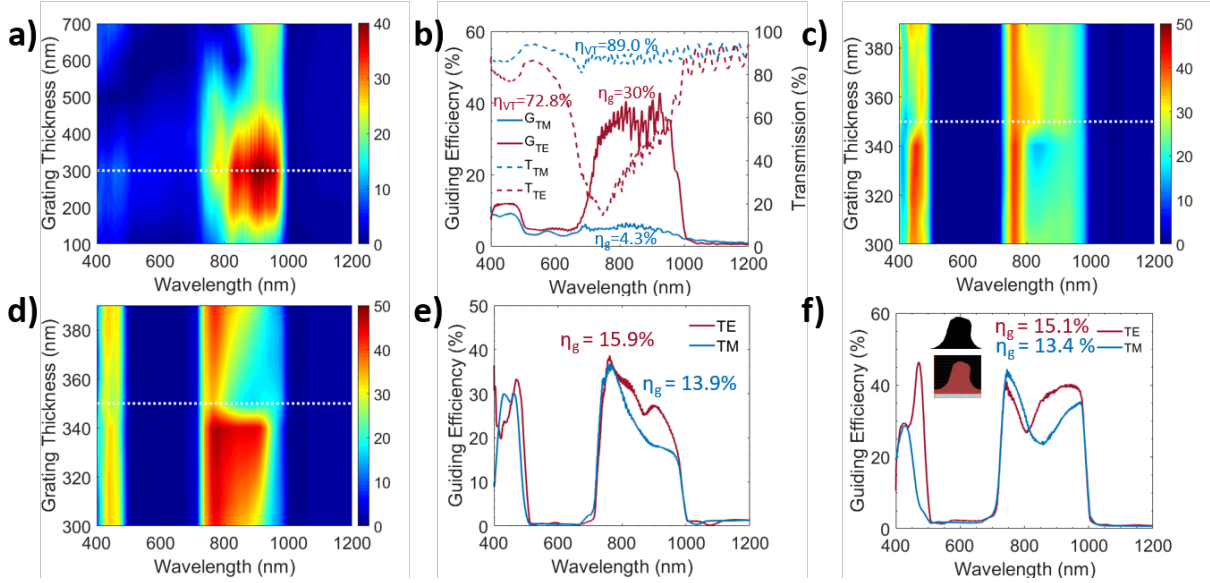

**Figure 2. Optimization of grating parameters using finite difference time domain analysis** a) TE incidence guiding efficiency spectrum as a function of grating thickness for Λ=680 nm & DC= 0.3. Similar studies are carried out for different duty cycles and DC= 0.3 gives better results. It can be inferred that for a grating thickness close to 300 nm (white dotted line at $t_g$=300 nm) the design shows better guiding. b) Optical guiding and optical transmission spectra of the planar concentrator for grating parameters Λ=680 nm, DC= 0.3, $t_i$= 0 nm & $t_g$= 300 nm for TE and TM incidences. The guiding efficiency ($\eta_g$) for TE and TM incidence is 30% and 4.3% respectively and averages to 16.2% in wavelength range of 700-1000 nm. The results show visible light transmission ($\eta_{vt}$) of 72.8% and 89.0% in the range of 400-700 nm for TE and TM incidences, respectively. c) & d) Polarization tolerant planar concentrator guiding spectra as a function of grating thickness for TE & TM respectively. At a grating thickness of 330 nm, the guiding efficiency spectra for both TE and TM closely matches e) TE & TM mode guiding efficiency spectra for grating parameters of Λ=660 nm, DC = 0.5, $t_i$= 90 nm & $t_g$= 340 nm, the guiding efficiency ($\eta_g$) for TE and TM are 13.9% and 12.5% in the wavelength range of 700-1000 nm respectively yielding an overall efficiency of 13.2%. f) Guiding efficiency spectra of the fabricated grating structure using the profile obtained from cross-sectional FIB/SEM is used to estimate the effect of non-rectangular profile and side wall roughness. The results shows a slight increase in guiding efficiency for TE incidence ($\eta_g$=15.1%) and slight decrease for TM ($\eta_g$=13.4%) incidence, reducing the overall efficiency. This decrease in efficiency can be due to the surface roughness and the change in profile, in ideal case the gratings are rectangular in shape with sharp edges, whereas the fabricated sample profile has round edges with rough surface. The inset shows the measured profile (top) and simulated profile (bottom) after profile extraction using image processing. The surface of the grating is considered as smooth in the ideal case but not in the fabricated sample. All these simulations are carried out using FDTD software from Lumerical.

Silicon nitride grating on glass can be tuned to be polarization dependent or independent light guiding. In our case, two basic designs are explored and simulated where guiding efficiency ($\eta_g$=30%), is maximized for TE incidence as shown in fig. 2(a). TE incidence guiding efficiency maximizes at a grating thickness of 300 nm. The grating period and duty cycle are kept constant at 680 nm and 0.3, respectively. TM incident guiding for the same design is minimal ($\eta_g$=4.3%), this can potentially find application in spectrally tuneable polarized beam splitter (fig 2(b)) etc. However, our design focuses on minimally polarization dependent structures for energy conversion using solar cells i.e. similar guiding spectra for both TE and TM polarized incidence. The spectral grating thickness shows a parameter space where TE (fig 2(c)) and TM (fig 2(d)) spectra show a matching guiding efficiency, with numerical values of 12.5% and 13.9% for TE and TM incidence respectively (averaged 13.2%) (fig 2 (e)). The optimized grating parameters are DC= 0.5, $t_i$= 90 nm, $t_g$= 340 nm & $\Lambda$=680 nm. However, we observe a slight variation in the fabricated structure compared to target parameter space (DC= 0.41, $t_i$= 130 nm, $t_g$= 270 nm & $\Lambda$=678 nm, fig. 3d). The grating profile is designed to have sharp rectangular edges but the measured profile using FIB/SEM shows round edge, which affects the spectral guiding efficiency moderately. The simulation results for the rounded edges show a small TM to TE polarization discrepancy of about 2% (fig 2(f)). For the simulations of measured profile, a duty cycle of 0.5 is considered as the grating widths are non-uniform across the 5mm sample and varies from 0.31 – 0.61 (fig. S3). The grating simulation parameters are $\Lambda$=680 nm, $t_g$= 395 nm, $t_i$ = 35 nm.

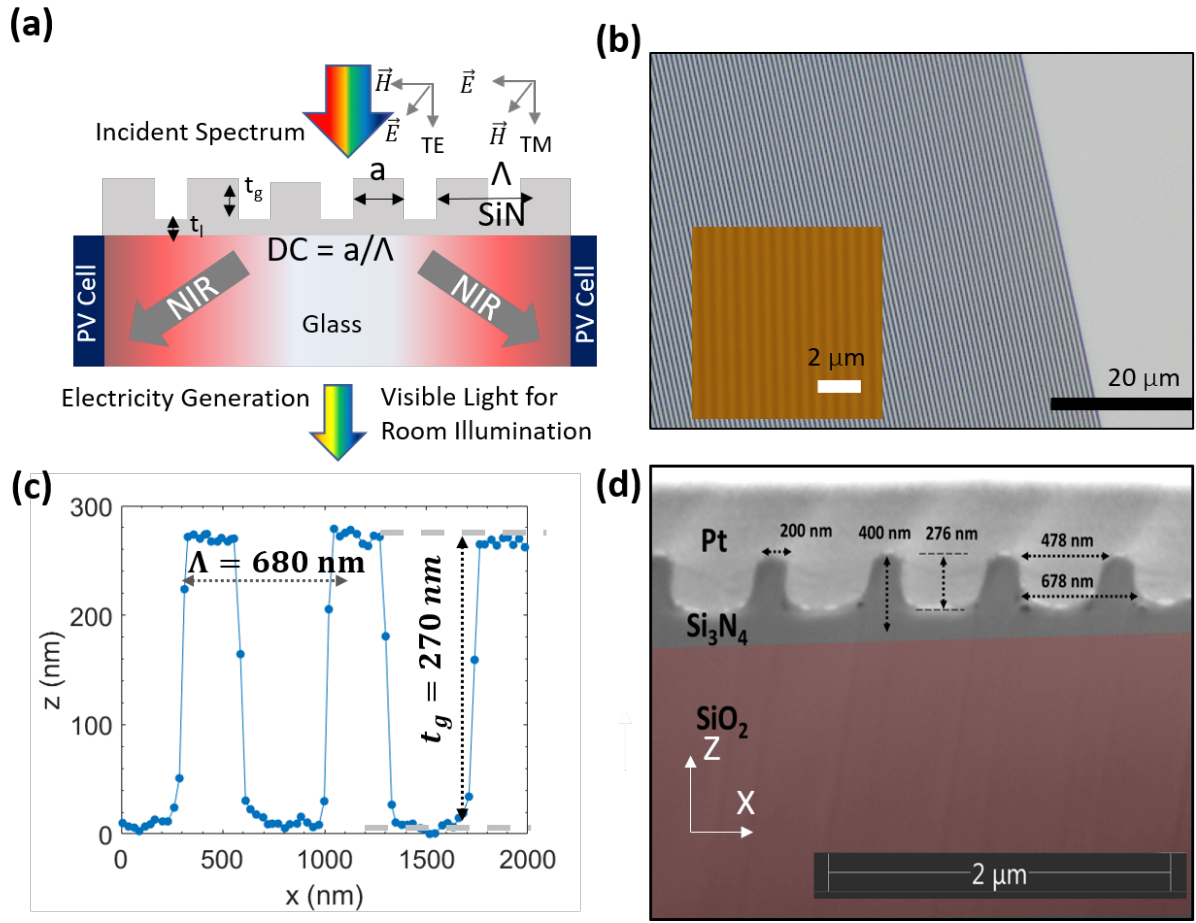

**Figure 3. Grating fabrication and characterization** a) Two-dimensional schematic of the PV cell integrated grating based planar concentrator. The gratings are shallow etched with silicon nitride layer on the glass substrate. The layer thickness is denoted as '$t_l$', grating thickness denoted as '$t_g$', the width of grating denoted as 'a', '$\Lambda$' is the grating period and 'a/$\Lambda$' is the grating duty cycle. b) Optical microscope image of e-beam fabricated grating at lower magnification and in the inset corresponding higher magnification image (yellow). Grating lines are fabricated over a large area of 5 mm x 5 mm using e-beam lithography and followed by reactive ion etching. c) Atomic force microscope image of the e-beam fabricated sample shows a grating thickness of 270 nm. d) Cross-sectional FIB/SEM image of the fabricated sample, the grating profile and the width is used in simulation to estimate the effect of sidewall roughness and profile variation due to the etching process.

The metasurface gratings are fabricated with e-beam lithography over a large area of 5 mm x 5 mm (fabrication process flow chart is given in fig. S1). The grating structures are relatively uniform over this area (fig. 3 (b)). Reactive ion etching is used as a pattern transfer into the silicon nitride using resist as a soft mask, and we obtain the height profile via atomic force microscope (3(c)). We find that the sidewalls are relatively straight, however, there is surface roughness on the grating surface with an RMS value of 5.78 nm calculated from the AFM (fig. S6). To gain further insights into the profile of the grating based metasurface, we performed cross-section focused ion beam milling (FIB) and imaged it with a scanning electron microscope (fig. 3(d)). Here, we find the grating profile with rounded edges as compared to the targeted sharp edges, which affects the guiding efficiency by a few percent, as discussed above (fig. 2(f)).

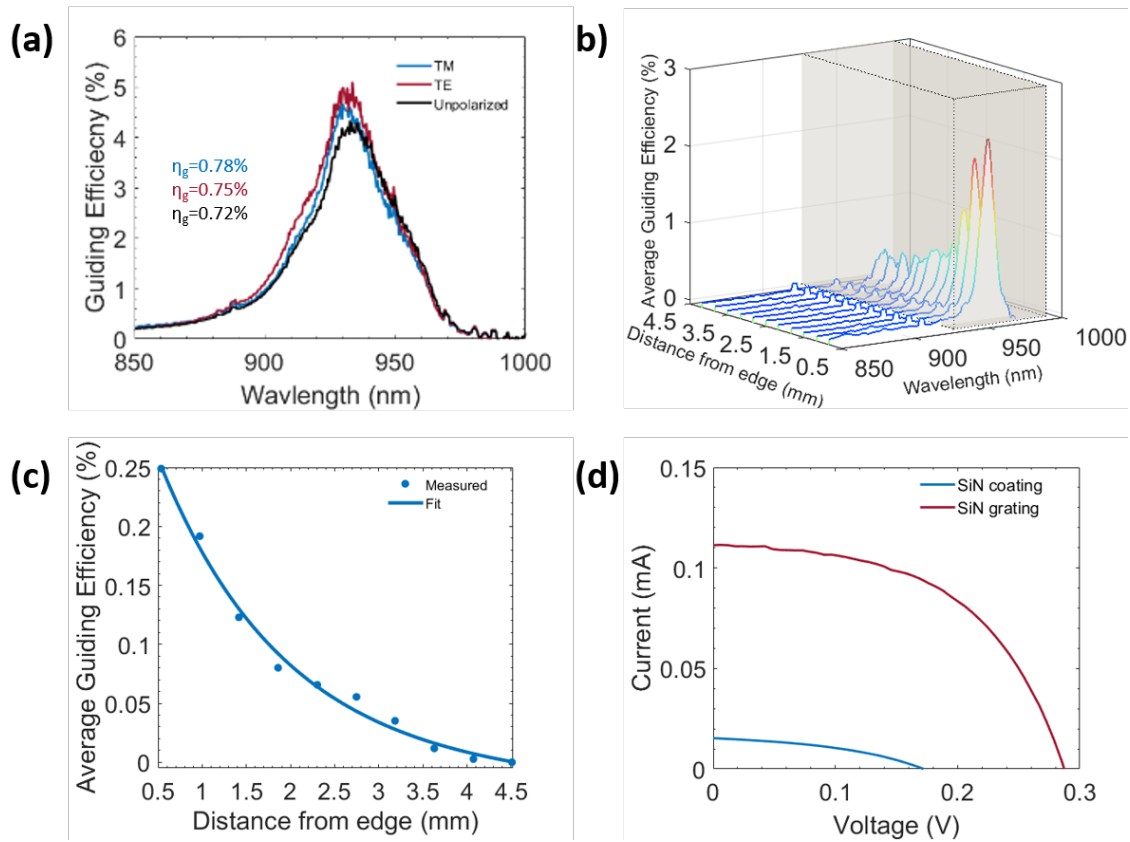

Figure 4. Optical characterization and performance evaluation of silicon nitride planar concentrator a) Guiding efficiency of the e-beam fabricated planar concentrator of 5 mm x 5 mm size. The beam spot size used in this measurement is ~4 mm in diameter so that the scattering from the edges can be avoided. The combined guiding efficiency for both the edges for TE, TM & unpolarized light are 0.78%, 0.75% & 0.72%, respectively in the wavelength range of 700-1000 nm b) Guided spectrum measured as a function of distance from the edge close to the integrating sphere, the guiding efficiency reduced with increase in distance from edge due to material absorption and scattering losses c) Average guiding efficiency as a function of distance from the end close to the collector (integrating sphere in this case). The spectra are obtained by measuring from one side by sweeping a beam (diameter of ~ 1 mm) along the length to evaluate the losses in the system. The exponential decay in the guiding efficiency suggesting a presence of both scattering and absorption loss in the system (dots show the experimental data points and solid line is the curve fit eqn. 3). d) I-V curve with attached solar cell on to the edge of the planar concentrator. Controlled sample is a glass coated with 400 nm thick silicon nitride and considered as the reference. Both the controlled sample and the grating sample of same area are tested under AAA class solar simulator with AM1.5G spectrum at 780 W/m$^2$. 11 times increase in short circuit current and 120 mV increment in open circuit voltage is observed for the grating integrated sample. These measurements are carried with a silicon solar cell attached to one edge and the efficiencies reported are doubled accordingly.

Testing the experimental implementation of the grating for power harvesting and for the multi-functional spectral properties discussed in fig. 1, here we show experiments demonstrating the concept functionality of the smart power window (fig. 4). We observe a peak in the guiding spectra at 935 nm (using measurement setup fig.S4), which is independent of polarization as designed (fig. 4(a)). The polarization independency requires two uncoupled guided waves to be simultaneously excited [44]. This excitation happens only for particular combinations of grating parameters [45]. Moreover, the experimental results for TE and TM incidence show similar guiding spectra (TE - 0.78%, TM - 0.75% & unpolarized - 0.72%). While these guiding efficiencies appear low as compared to the numerical results, they can be attributed to various losses in the system including material absorption, optical guiding loss, scattering loss due to roughness and non-uniformity of duty cycle, grating thickness across the sample. Even when there is low optical absorption in the substrate, effects such as scattering accumulate for longer optical path lengths. The glass substrate acts as a multimode waveguide with different modes propagating at different angles (eqn. 1) and results in different optical paths (i.e. varying k-vectors). For the guided modes, the optical path is longer for lower wavelengths and vice versa. Additionally, surface roughness in optical materials may cause additional optical scattering losses in a propagating mode. This may be due to defects and impurities in the substrate.

We use the cutback technique known from optoelectronics to investigate the effective optical loss (Fig. 4(b)&(c)). For the effective loss evaluation, the guiding efficiency is measured as a function of distance of beam incidence position from the edge of the substrate facing the integrating sphere. Indeed, we find a declining efficiency with distance to the sample edge. The efficiency is decreased due the losses in the material. This can be understood from a multitude of effects such as material absorption, scattering and diffraction at silicon nitride layer/grating interface. Material losses play a major role in the performance of the planar concentrator. We attempt to model the losses in the system assuming that there are mainly two components a linear- and an exponential loss component (Eqn. 3). The linear component is a due to scattering centers due to defects and impurities of the substrate and the exponential component as result of material absorption. From Beer- Lambert's law, it is understood that the absorption has an exponential relation with the length. The guiding efficiency as function of distance to the solar concentration edge as shown in Fig. 4(c) and the measured data points are curve with eqn. 3 to obtain a model to estimate the losses.

$$g(x) = c_1 e^{-c_2 x} - c_3 x \qquad (3)$$

Where, $c_1$=0.37, $c_2$=0.72 & $c_3$=0.003.

Fitting our distance-related optical powers at the sample edge with eqn. 3 (i.e. only counting absorption and scattering), we find a reduction of about 87% of the initial optical power due to losses (average propagation length = 2.5 mm). Thus, if all losses are eliminated the guiding efficiency would be 7X of the experimentally measured value i.e. 5.7% (Fig. 4(a)). The main component contributing to loss is being the absorption, for every 1.4 mm the loss becomes 1/e, i.e. the attenuation length. The contribution from scattering is linear and minimal compared to absorption i.e. for the same 1.4 mm distance the loss due to scattering is only 0.02% of total loss. However, scattered light can also contribute to the absorption loss since it is guided in the substrate by increased path length.

Next, we test the smart power window prototype system performance by integrating a silicon solar cell at the edge of the grating sample capturing the guided spectral power. The I-V characteristics of solar cell (4(d)) shows more than 10 times increment in the short circuit current and 120 mV increase in open circuit voltage (table 1) with the SiN grating based metasurface as compared to a controlled sample of SiN layer on glass with 400 nm thickness (the same thickness of silicon nitride used in the grating structure). The solar cell is cut into an area of 16 mm$^2$ and covered with a copper sheet coated with black coating to make it to an area of 7 mm$^2$ to match the edge area of the sample. We note, that this will introduce additional parasitic shading losses from the solar cell, which may adversely affect the performance of the system. The integrated system yields a conversion efficiency of 4.2% (assuming no losses) under the solar simulator with an irradiance of 780 W/m$^2$. The solar cell converts almost 80% percent of the incoming radiation, as there is lower thermalization loss in the guided spectrum.

Table 1: Solar cell parameters obtained with solar simulator under AM1.5G illumination with 780 W/m$^2$ for cell place only at one edge.

| | Optical guiding Efficiency (%) | $V_{oc}$(V) | $I_{sc}$ (mA) | $P_{max}$ (mW) | FF(%) | Electrical Efficiency (%) |
|---|---|---|---|---|---|---|
| Control Sample (e.g. SiN) | <0.01 | 0.17 | 0.01 | <0.01 | 40 | 0.01 |
| Smart Power Window | 0.75 | 0.29 | 0.11 | 0.02 | 52 | 0.65 |

## Conclusion

We demonstrated a transparent planar concentrator based on a dielectric grating based metasurface integrated onto a glass substrate. This system has a unique potential in generating electricity from NIR portion of the sunlight incident on the windows of modern buildings. The dielectric metasurface grating functions as a spectral filter on the incoming sunlight, which allows visible light to pass through the glass windows but guides the NIR light towards the glass edge for efficient PV conversion. We have demonstrated a prototype of dimension 5 mm x 5 mm using e-beam lithography. The integrated device shows an optical guiding efficiency of 5.25% and solar cell conversion efficiency of 4.2% in the range of 700-1000 nm (considering no loss case) with visible light transmission of 72% and 88%. As such, this planar concentrator system is well suited for building integrated photovoltaics application and can become an integral part of future on-site electricity generation, heating and cooling infrastructure for modern buildings. Such device in principle can be fabricated on a large area using cost-effective techniques such as Interference Lithography, stepper lithography and Nano-Imprint Lithography.

## Methods

### Design and Simulation

Rigorous Coupled Wave Analysis (RCWA) based software from GSolver© is used to understand the diffraction efficiencies of various material combinations and a preliminary tool for choosing material. RCWA considers an infinitely periodic structure and yields only the diffraction efficiencies of various diffraction orders. Finite Difference Time Domain (FDTD) simulation methodology from Lumerical is used for design of a finite device and to obtain guiding at the edge of the substrate. The optimization process aims at maximizing the guiding efficiency followed by minimizing the polarization dependency. Simulations are done using a non-uniform mesh with mesh accuracy of 2. All the simulations consider two-dimensional geometry with PML and symmetric boundary conditions. The material refractive index for silicon nitride utilizes the ellipsometer measured data (fig. S2) and the substrate data is from Palik database.

### Fabrication

We developed a large area Electron Beam Lithography (EBL) process to fabricate silicon nitride grating on glass. After cleaning the glass substrate, it is coated with 400 nm of PECVD silicon nitride. Silicon nitride deposition is characterized using dektak profilometer and ellipsometry for thickness and refractive index control. Since the photoresist used for the EBL step (PMMA) is thin and shows poor selectivity to the dry etch process of silicon nitride, instead of using the resist itself as a mask, we deposit an additional 30 nm thin chromium layer on the wafer as the etch mask. The optimized device geometry design (Fig. 2) is then used for the fabrication of the grating prototype. Using the EBL exposure, an area of 5 mm x 5 mm grating is fabricated on a 2 mm thick glass substrate. PMMA is spin coated at an rpm of 4000 for 45 seconds followed by baking it at $180°$ C for 2 minutes. The EBL tool uses an electron high tension of 30 keV with an aperture of 30 $\mu$m at a dosage of 400 $\mu C/cm^2$. The chromium layer is then wet etched with PMMA as masking layer using CR7. After removal of PMMA with acetone, RIE etch is carried out to etch the grating features on silicon nitride. The etch depth is characterized by AFM and FIBSEM (Fig. 3).

Optical Characterization:

The optical measurements are carried out to obtain transmission, reflection and guiding efficiency. The results for e-beam sample are shown in fig. 4. The setup uses a fibre coupled halogen lamp from Ocean optics HL 2000 followed by collimating optics with chromatic aberration correction. The light guided through the glass substrate is collected using an integrating sphere (ISP-R from Labsphere) and fed to the spectrometer (Ocean optics USB2000).

Solar cell IV characterization is carried out using Sol-3A solar simulator from Oriel Instruments at AM1.5 illumination of 780 $W/m^2$. The IV characteristics are measured using a Keithley 2440 source meter.

Supporting Information. A listing of the contents of each file supplied as Supporting Information should be included. For instructions on what should be included in the Supporting Information as well as how to prepare this material for publications, refer to the journal's Instructions for Authors.

AUTHOR INFORMATION

Corresponding Author

Dr Bala Pesala, Principal Scientist, CSIR- Central Electronics Engineering Research Institute, Chennai, India

**Present Addresses**

†If an author's address is different than the one given in the affiliation line, this information may be included here.

**Author Contributions**

The manuscript was written through contributions of all authors. All authors have given approval to the final version of the manuscript. ‡These authors contributed equally.

**ACKNOWLEDGMENT**

The author AE would like to acknowledge IUSSTF for providing BASE fellowship.

(17) Karp, J. H.; Tremblay, E. J.; Ford, J. E. Planar Micro-Optic Solar Concentrator. *Opt. Express* 2010, *18* (2), 1122–1133.

(18) Baker, K.; Karp, J.; Hallas, J.; Ford, J. Practical Implementation of a Planar Micro-Optic Solar Concentrator. In *Proceedings of SPIE - The International Society for Optical Engineering*; 2012; Vol. 8485, p 848504.

(19) Karp, J. H.; Tremblay, E. J.; Ford, J. E.; Van Sark, W. G.; Barnham, K. W.; Slooff, L. H.; Chatten, A. J.; Büchtemann, A.; Meyer, A.; McCormack, S. J.; et al. Planar Micro-Optic Solar Concentrator. *Sol. Energy Opt. Lett. Handb. Opt. M. H. Chou, M. Appl. Phys. Quantum Electron. J. Micromech. Microeng* 2004, *705* (73), 423–430.

(20) Meinardi, F.; Bruni, F.; Brovelli, S. Luminescent Solar Concentrators for Building-Integrated Photovoltaics. *Nat. Rev. Mater.* 2017, *2*, 1–9.

(21) Li, H.; Wu, K.; Lim, J.; Song, H. J.; Klimov, V. I. Doctor-Blade Deposition of Quantum Dots onto Standard Window Glass for Low-Loss Large-Area Luminescent Solar Concentrators. *Nat. Energy* 2016, *1* (12).

(22) Tosun, T. Luminescent Solar Concentrator. 2015, *2* (1), 21–23.

(23) Desmet, L.; Ras, A. J. M.; de Boer, D. K. G.; Debije, M. G. Monocrystalline Silicon Photovoltaic Luminescent Solar Concentrator with 4.2% Power Conversion Efficiency. *Opt. Lett.* 2012, *37* (15), 3087–3089.

(24) Banal, J. L.; Ghiggino, K. P.; Wong, W. W. H. Efficient Light Harvesting of a Luminescent Solar Concentrator Using Excitation Energy Transfer from an Aggregation-Induced Emitter. *Phys. Chem. Chem. Phys.* 2014, *16* (46), 25358–25363.

(25) Wang, T.; Zhang, J.; Ma, W.; Luo, Y.; Wang, L.; Hu, Z.; Wu, W.; Wang, X.; Zou, G.; Zhang, Q. Luminescent Solar Concentrator Employing Rare Earth Complex with Zero Self-Absorption Loss. *Sol. Energy* 2011, *85* (11), 2571–2579.

(26) Kostuk, R. K.; Castillo, J.; Russo, J. M.; Rosenberg, G. Spectral-Shifting and Holographic Planar Concentrators for Use with Photovoltaic Solar Cells. *Proc. SPIE* 2007, *6649*, 66490I–66490I–8.
20